\documentclass[12pt]{article}

\usepackage{graphicx}
\usepackage{psfrag}   
\newcommand{\qq}{{\overline q}q}
\newcommand{\uu}{${\overline u}u \ $}
\newcommand{\dd}{${\overline d}d \ $}

\newcommand{\KK}{{\overline K}K}
\newcommand{\st}{${\overline s}s \ $}
\newcommand{\be}{\begin{equation}}
\newcommand{\ee}{\end{equation}}

\begin{document}
\large
\hfill\vbox{\hbox{DCPT/07/22}
            \hbox{IPPP/07/11}}
\nopagebreak

\vspace{2.0cm}
\begin{center}
\LARGE
{\bf{Can experiment distinguish tetraquark scalars, molecules and $\qq$ mesons?}}

\vspace{12mm}

\large{M.R. Pennington}

\vspace{5mm}
{\it Institute for Particle Physics Phenomenology,\\ Durham University, 
Durham, DH1 3LE, U.K.}
\end{center}
\vspace{1cm}

\centerline{\bf Abstract}

\vspace{3mm}
\small
Scalar mesons are a key expression of the strong coupling regime
of QCD. How those with $I_3=0$ couple to two photons supplies important 
information about which is transiently ${\overline q}q$ or 
${\overline {qq}} qq$ or  multi-meson molecule or largely glue, provided
(i) we know the location of the corresponding poles in the complex energy plane, and
(ii) we have reliable predictions from strong coupling QCD of the radiative widths for these compositions.
Number (i) is being supplied by careful analyses of high statistics data 
particularly that coming from  $B$, $D$, $J/\psi$ and $\phi$ decays. Number (ii) is still required, before we can answer the
question in the title. 

\normalsize

\newpage

\parskip=2.mm
\baselineskip=5.8mm

\section{QCD vacuum and the role of scalars}

\noindent 
Like many questions, that in the title on the nature of the scalar mesons prompts further questions, some of which can be answered in models as will be discussed.
However, first let us ask: why should we care what scalars are made of? The answer is that scalars are special. They constitute the Higgs sector of the strong interaction giving masses to all of the light hadrons. They are in turn intimately tied to the structure of the QCD vacuum. To understand this we have to go back 60 years.
We know that just counting quarks is sufficient to explain why a typical meson, like the $\rho$, and the lightest baryon, the nucleon, have masses in the ratio of $\,2:3$. Then the question is why are pions so very light? Indeed, now that we know that the current masses of the {\it up} and {\it down} quarks are just a few MeV, one could equally ask why are the masses of typical mesons and baryons so heavy!
The answer is in the structure of the QCD vacuum. While the vacuum of QED is essentially empty with just perturbatively calculable corrections 
from virtual photon emission and absorption (including electron-positron loops), that of QCD is quite different. While asymptotic freedom ensures that quarks propagate freely over very short distances deep inside a hadron, little perturbed by the sea of quark-antiquark pairs and the clouds of gluons, over larger distances (distances of the order of a fermi), the interactions between quarks and gluons become so strong that they produce correlations in the vacuum. Correlations that we know as condensates. It is by moving through this complex vacuum that light quarks gain their mass.

We are of course familiar with the idea that the world of hadron physics reflects the symmetry properties of the underlying world of quarks and gluons.
The hadron world has isospin symmetry because of the near equal mass of the {\it up} and {\it down} quarks and their flavour blind interactions. The fact that the current masses of these quarks are so much lighter than the natural scale of QCD, {\it viz.} $\Lambda_{QCD}$, means that we can regard them as massless. Then their helicity becomes a well-defined quantum number and the left and right-handed worlds decouple. QCD has an almost exact $SU(2) \times SU(2)$ chiral symmetry. In the world of hadrons this symmetry is not apparent. Scalars and pseudoscalars, vectors and axial-vectors are not degenerate in mass with closely related interactions. Long long ago before QCD was uncovered, Nambu proposed a mechanism for this spontaneous breakdown of chiral symmetry~\cite{nambu}. If the potential generated by the interactions of a scalar ($\sigma$) field and the pseudoscalar ($\pi$) has a Mexican hat shape, nature can choose a ground state in which
the $\sigma$ has a non-zero vacuum expectation value. Then pions, which correspond to the quantum fluctuations round the bottom of the hat, are massless. In contrast, the physical scalar field, produced  by fluctuations up and down the sides of the hat,  make it massive. In the underlying world of QCD, it is the operators $\,\qq$, ${\overline q} G q$, {\it etc.}, that gain non-zero vacuum expectation values. To realise 
Nambu's picture, this dynamical breaking of chiral symmetry must be dominated by the $\qq$ condensate.

This is all very well-known. What is relatively new, is that these ideas have been tested in experiment and found to agree exactly with this picture. For twenty years QCD sum-rules have indicated that the value of $\,\langle\ \qq \, \rangle \simeq -(240$ MeV$)^3$ and it is this 240 MeV that sets the scale for the constituent mass of the {\it up} and {\it down} quarks.
Similarly, Schwinger-Dyson calculations of the Euclidean momentum dependence of the quark mass function show that such a value for the chiral condensate follows just from $\Lambda_{QCD}$. Moreover, because pions are the Goldstone bosons of chiral symmtery breaking, their interactions at low energy directly reflect the value of this condensate. 
These interactions can be studied experimentally, for instance in $K_{e4}$  
decay. There a $K$ turns into an electron and its neutrino, plus two pions, which once formed interact strongly. Because pions are the lightest of all hadrons, their interactions are universal. Thus from the $K_{e4}$ decay distribution we can learn about the phase of the relevant $\pi\pi$ interaction. The first high statistics measurement came 5 years ago from E865 at BNL~\cite{e865}, this showed that the low energy behaviour of the phase-shift corresponded to a $\qq$ condensate of exactly the size expected~\cite{colangelo}. This scale for the $\qq$ condensate has been further checked in the last few months by the preliminary results from NA48~\cite{na48}.

Thus we see that Nambu's picture of the breaking of chiral symmetry is indeed realised in nature. So  what is the scalar field,
the non-zero vacuum expectation value of which generates masses for all light hadrons?  Do we identify Nambu's $\sigma$ with the lightest $I=J=0$ state, the $f_0(600)$~? That such a meson might exist was proposed more than 40 years ago. However, 
Breit-Wigner fits to $\pi\pi$ mass distributions have found a wide range of masses and widths $\simeq {\cal O}(1$ GeV) over the years, so short-lived that it was far from certain that it really existed. Now we know, thanks to a rather precise analytic continuation into the complex energy plane, that a pole does exist, signalling a state in the spectrum of hadrons. But is this state the chiral partner of the $\pi$? Is it the Higgs of the strong interaction? There exist many other isoscalar scalars~\cite{pdg} $f_0(980), f_0(1370), f_0(1510), f_0(1720)$. What role do these play? What is more, these have isodoublet and isotriplet partners, like the $K_0^*(800)$, $K_0^*(1430)$, $a_0(980)$ and $a_0(1430)$. How are these all related? The quark model would lead us to expect a $\,^3P_0\,$ nonet, but there are 19 scalars (counting all the different charged states) just listed. Enough for two nonets and  have an isosinglet left over! Of course, there is a danger of subjectivity in what is included amongst this list. That the nine lightest do not fit a simple $\qq$ pattern anticipated from vector and tensor mesons is readily seen by noting that
both the $f_0(980)$ and $a_0(980)$ sit very close to the $\KK$ threshold and couple very strongly to these channels. In a simple quark multiplet only the \st state does that~\cite{morgan-weinstein}.
\begin{figure}[t]
\begin{center}
\includegraphics[width=12.cm]{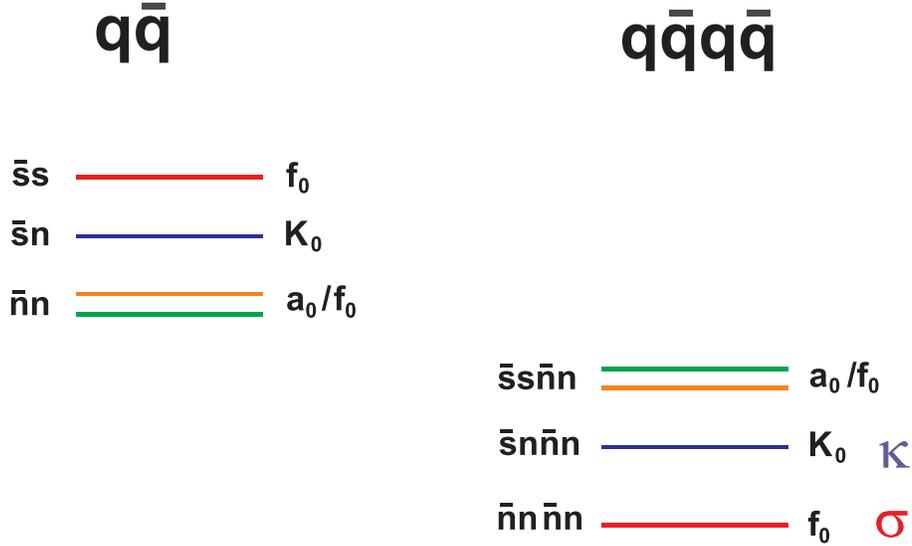}
\vspace{2mm}
\caption
{The spectrum of scalar states given by a simple {\it ideal} nonet of ${\overline {q}}q$ and of ${\overline{qq}qq}$ mesons, where $n = u,d$, indicating which would be identified with the isoscalar $\sigma$ and its isodoublet partner, the $\kappa$, if the observed hadrons had these presumed {\it orthogonal} compositions.} 
\end{center}
\end{figure}

Now at this meeting we have heard a lot about tetraquark mesons~\cite{jaffe,schechter,maiani}, where colour $\overline 3$ scalar diquarks, formed from quarks of different flavour, like $[ud]$, $[us]$ and $[ds]$, bind with the corresponding anti-diquarks. The lightest multiplet is a nonet and lighter~\cite{jaffe} than the corresponding $^3P_0$ $\qq$ nonet, as in Fig.~1. Moreover, the heaviest tetraquarks  are an $f_0$ and $a_0$: ${\overline {[ns]}}[ns]$, with $n=u,d$, suggesting these might be good assignments for the $980$ MeV states. However, Weinstein and Isgur~\cite{wi} found within their potential model that the only scalar tetraquark systems to bind would be in the form of  $\KK$-molecules. This would predict just four states, the $f_0$ and $a_0$, and not a complete nonet.   In the theorists' favourite large $N_c$ limit, the $\qq$ multiplet becomes stable, while the tetraquark mesons merge into the continuum. However, we do not live in a world in which $N_c$, the number of colours, is large. Indeed, as we will comment later, nowhere is this more apparent than in the scalar sector. With the addition of a glueball with mass computed on the lattice~\cite{latticeglueball} in a world without quarks to be $1.5-1.7$ GeV, we have our 19 \lq\lq lightest'' scalars. Of course, these states $\qq$, ${\overline {qq}}qq$ (or $\KK$-molecules) and glueball are not orthogonal to each other, they inevitably mix~\cite{schechter,closemix}.
So how can we try to distinguish which scalars have what composition?

\section{Two photon interactions: shedding light on scalars}

Two photon interactions can illuminate these issues.
For orientation, let us presume we have extracted the two photon couplings of resonant states. Before looking at the enigmatic scalars, let us consider 
 the two photon couplings of the tensor mesons, $f_2$, $a_2$ and $f_2'$, that belong to an ideally mixed quark multiplet. Their radiative widths are proportional to the square of the average charge squared of their constituents, Fig.~2.
The absolute scale of their couplings depends on dynamics: on how the ${\overline q}q$ pair form the hadron. In the non-relativistic limit, as with charmonia, this is simply related to the wavefunction at the origin. If we assume that these are equal for the tensor states then we have the prediction that
\begin{figure}[t]
\begin{center}
\includegraphics[width=6.cm]{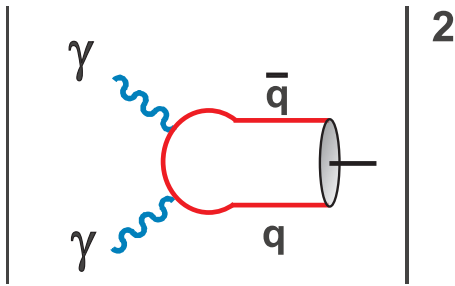}
%\vspace{2mm}
\caption
{Two photon decay rate of a hadron is the modulus squared of the amplitude for $\,\gamma\gamma\,$ to produce a $\,{\overline q}q\,$ pair and for these to bind by strong coupling dynamics. Present calculations use crude approximations. A genuine relativistic strong coupling approach is required, particularly for the lightest pseudoscalars and scalars. } 
\end{center}
\end{figure}
\be
\Gamma(f_2\to\gamma\gamma)\;:\;\Gamma(a_2\to\gamma\gamma)\;:\;\Gamma(f_2'\to\gamma\gamma)\;=\; 25\,:\,9\,:\,2\quad .
\ee
Experiment~\cite{pdg} is in reasonable agreement with this, given the uncertainties in extracting  couplings from data (as we shall see).
If we apply the same ideas to the pseudoscalars, $\pi^0, \eta, \eta'$, though not ideally mixed, we would not be able to reproduce experiment~\cite{pdg}:
\be
\Gamma(\pi^0\to\gamma\gamma)\;:\;\Gamma(\eta\to\gamma\gamma)\;:\;\Gamma(\eta'\to\gamma\gamma)\;\simeq\; 1\,:\,60\,:\,500\quad .
\ee
While the non-relativistic quark model  determines the intrinsic coupling
to photons, it does not include the dynamics of binding the quark and antiquark into a meson, Fig.~2. Gauge invariance of a Lagrangian for the  pseudoscalar-photon-photon interaction would introduce a factor of mass cubed for each meson. As noted by Hayne and Isgur~\cite{hayneisgur}, such factors are just right to bring the quark model prediction into agreement with experiment of Eq.~2. However, what this indicates is that we should not be using the non-relativistic quark model and correcting it for light quarks. Rather we need genuinely relativistic strong coupling calculations of such radiative widths. This we will see equally applies to the scalars.
\begin{figure}[t]
\begin{center}
\includegraphics[width=12.5cm]{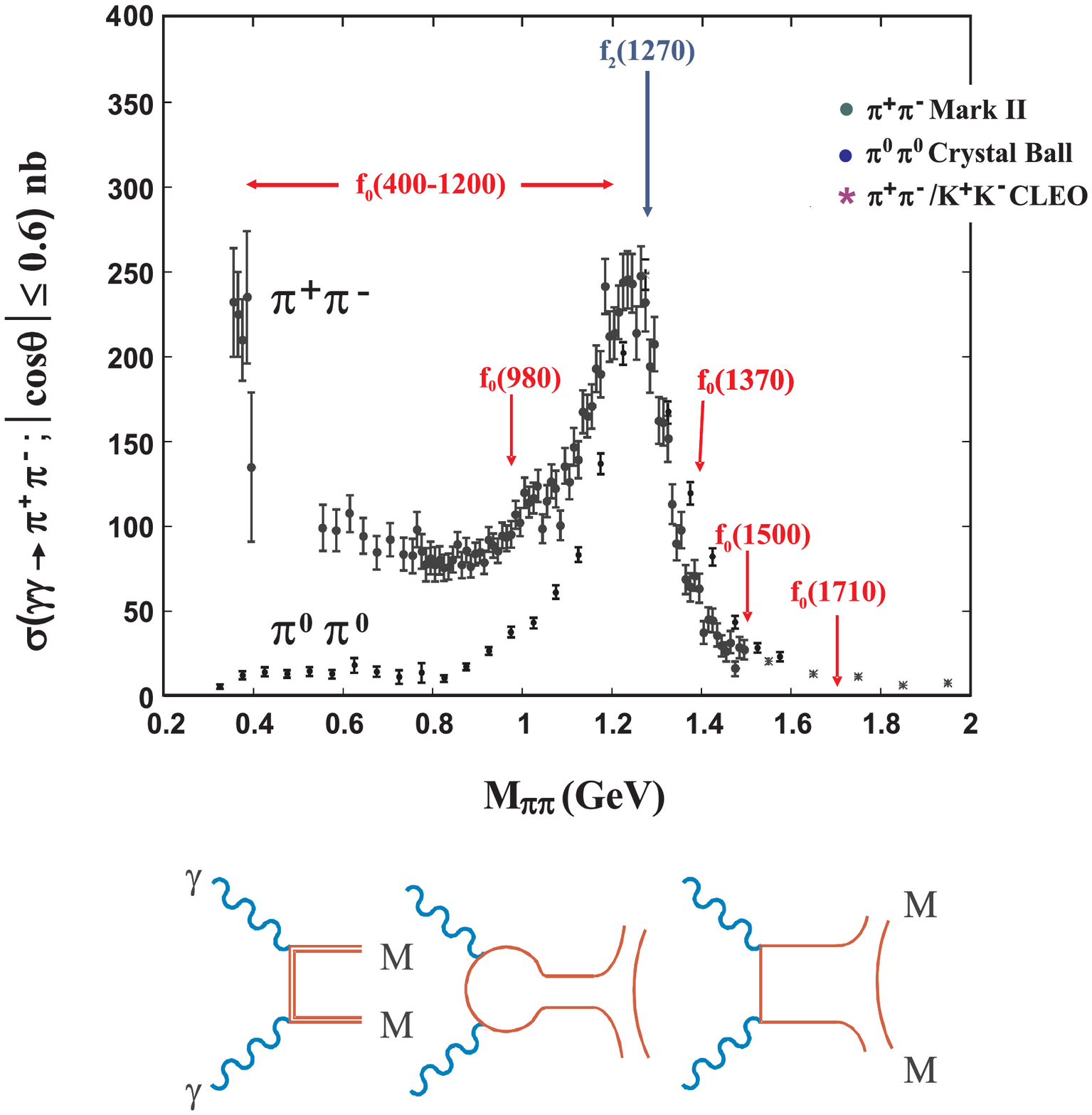}
\caption
{Integrated cross-section for $\gamma\gamma\to\pi\pi$ as a function of c.m. energy $M(\pi\pi)$ from Mark II~\protect\cite{MarkII}, Crystal Ball (CB)~\protect\cite{CB} and CLEO~\protect\cite{cleo}. The $\pi^0\pi^0$ cross-section has been scaled to the same angular range as the charged data and by an isospin factor for the $f_2(1270)$ peak. Below are graphs describing the dominant dynamics in each kinematic region,
as discussed in the text. } 
\end{center}
\vspace{-3.5mm}
\end{figure}
\begin{figure}[t]
\begin{center}
\includegraphics[width=9.5cm]{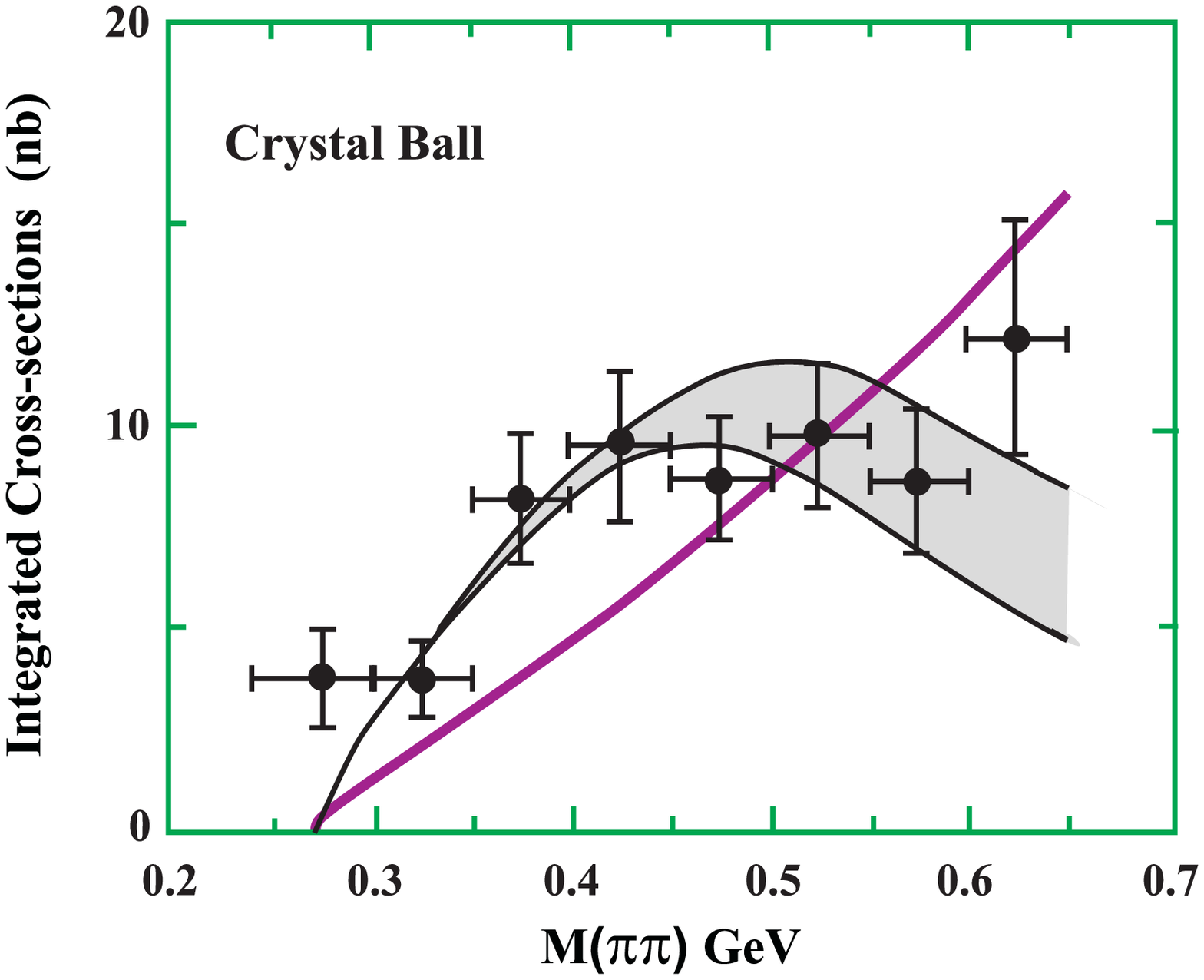}
\caption
{Cross-section for $\gamma\gamma\to\pi^0\pi^0$ integrated over $|\cos \theta^*| \le 0.8$ as a function of $M(\pi\pi)$. The data are from Crystal Ball~\protect\cite{CB}. The line is the prediction of $\chi PT$ at one loop $(1\ell)$~\protect\cite{bijnens}. The shaded band shows the dispersive prediction~\protect\cite{mp,mrpgamma} --- its width reflects the uncertainties in 
experimental knowledge of both $\pi\pi$ scattering and vector exchanges.} 
\end{center} 
\end{figure}

Now how do we extract such two photon couplings?  Let us consider the exclusive process
for which we have the most information, {\it viz.} $\gamma\gamma\to\pi\pi$, with both charged and neutral pions~\cite{MarkII,CB,cleo}, as displayed in Fig.~3. At low energy, the photon has long wavelength and interacts with the whole of the hadron. Consequently, it \lq\lq sees'' the charged pions, but not the neutral. The $\pi^0\pi^0$ production is small; $\pi^+\pi^-$ is large (Fig.~3). How large is determined by the charge of the pion. However, as the energy increases (by
just 1 GeV) the wavelength of the photon shortens sufficiently that it recognises that pions, whether charged or neutral, are made of the same charged constituents, namely quarks, and causes these to resonate. It produces the prominent tensor meson $f_2(1270)$, as well as the scalars which are more difficult to discern. Then as the energy increases further, the photon probes not constituent quarks, but current quarks. The interactions of these \lq\lq undressed'' quarks can be calculated perturbatively as shown by Brodsky and Lepage~\cite{brodsky-lepage}. If one allows for an adjustable normalisation, these predictions for both the energy and c.m angular dependence compare well with the latest data from Belle~\cite{BelleMM}.
What we are interested in here is the scalar signal in Fig.~3 underneath the large $f_2(1270)$ from $\pi\pi$ threshold upwards.

To see how to extract this signal, let us focus on $\pi\pi$ production close to threshold. Since pions are the Goldstone bosons of chiral symmetry  breaking, their interactions at low energy can be computed (as already remarked for $K_{e4}$ decays) as a function of momentum. For $\gamma\gamma\to\pi^0\pi^0$, the lowest order in chiral perturbation theory ($\chi PT$) involves one loop graphs which predict a cross-section rising almost linearly with energy from threshold~\cite{bijnens}, as seen in Fig.~4. This is in rather poor agreement with the only normalised data from Crystal Ball~\cite{CB}, shown already in a bigger energy range in Fig.~3. This was thought to be a little worrying, since Maiani~\cite{maianichpt} described this process as providing a \lq\lq gold-plated'' prediction of $\chi PT$, suggesting that the data had to  be wrong. However, one can calculate this cross-section almost exactly at low energy, by noting that
 neutral pions can be produced by first $\gamma\gamma\to\pi^+\pi^-$ by the Born amplitude and then $\pi^+\pi^-$ can interact to form the $\pi^0\pi^0$ final state. That this cross-section is calculable from strong interaction dynamics makes it almost unique among hadronic processes. This is because of Low's low energy theorem~\cite{low,mrpgamma}. This states that the amplitude for $\gamma\pi\to\gamma\pi$ at threshold is exactly given by the one pion exchange Born amplitude. Though this applies at just one kinematic point, the amplitude smoothly approaches this limit throughout the low energy region. This is because of the closeness of the $t$ and $u$-channel pion poles to the three physical regions. This means that the Born amplitude modified by final state interactions controls the $\gamma\gamma\to\pi\pi$ amplitude, as long as the pion pole is  much closer than any other exchange: the next nearest being the $\rho$ and~$\omega$.
Consequently, if we consider the analytic structure of the partial wave amplitudes, they have a left hand cut generated by crossed channel exchanges, in which essentially the part from $s=0$ to $s\simeq -m_{\rho}^2\,$ is known from pion exchange. Their right hand cut is controlled by direct channel dynamics and the discontinutity across this cut is specified by unitarity:
\be
{\rm Im}\, {\cal F}(\gamma\gamma\to\pi\pi;s)\;=\;\sum_n\;{\cal F}(\gamma\gamma\to H_n;s)^*\,{\cal T}(H_n\to\pi\pi;s)\quad ,
\ee
where  ${\cal F}$, ${\cal T}$ are partial wave amplitudes with definite isospin
and the sum is over all allowed intermediate states $H_n$.  In the low energy region, only the $\pi\pi$ intermediate state is kinematically possible. Consequently, knowing the nearby left hand cut discontinutity from $\pi$-exchange and the right hand cut discontinuity from Eq.~(3), one can compute~\cite{mp} the $\gamma\gamma\to\pi\pi$ partial waves accurately up to $\sqrt{s} \equiv M(\pi\pi) \simeq 550$ MeV. The experimental inputs are the $\pi\pi\to\pi\pi$ partial waves. These give~\cite{mrpgamma,mrp-prl} the cross-sections for $\gamma\gamma\to\pi^0\pi^0$
 and $\pi^+\pi^-$ in agreement with experiment from Crystal Ball~\cite{CB} and Mark II~\cite{MarkII} respectively. That for $\pi^0\pi^0$ is shown as the band in Fig.~4.

The $\pi\pi$ partial waves from experiment, of course, reflect not just the lowest order in $\chi PT$ but all orders. This suggests that higher orders in $\chi PT$ must substantially correct the one loop line in Fig.~4. Not surprisingly subsequent two loop $\chi PT$ calculations brought better agreement~\cite{bellucci} with experiment.
That the $\gamma\gamma\to\pi\pi$ amplitudes are reliably calculable at low energy from the Born amplitude modified by final state interactions has two important applications, to which we now turn.

\section{Two photon coupling of the $\sigma$}

A key application of dispersion relations to $\pi\pi$ scattering is in locating the $\sigma$-pole in the complex energy plane. As always in a fixed-$t$ dispersion relation the left hand cut requires knowledge of crossed-channel exchanges. The unique feature of $\pi\pi$ scattering is its 3-channel crossing symmetry. This means both the right and left hand cuts involve integrals over $\pi\pi\to\pi\pi$ scattering.
Making use of this crossing property, and then partial wave projecting, leads to the Roy equations. Solving this set of rigorous equations inputting experimental information above 800 MeV, together with chiral constraints, allows the $I=J=0$ $\pi\pi$ partial wave to be determined everywhere on the first sheet of the energy plane, Fig.~5.
As shown by Caprini, Colangelo and Leutwyler~\cite{ccl}, this  fixes a zero of the $S$-matrix at $E\,=\,441\,-i\,227$ MeV, which reflects a pole on the second sheet at the same position.
 This not only confirms the $\sigma$ as a state in the spectrum of hadrons, but locates the position of its pole very precisely with errors of only $\sim \pm 20$ MeV. This is in agreement with the related analysis by Zhou {\it et al.}~\cite{zhou}.
\begin{figure}[hb]
\begin{center}
\includegraphics[width=12.5cm]{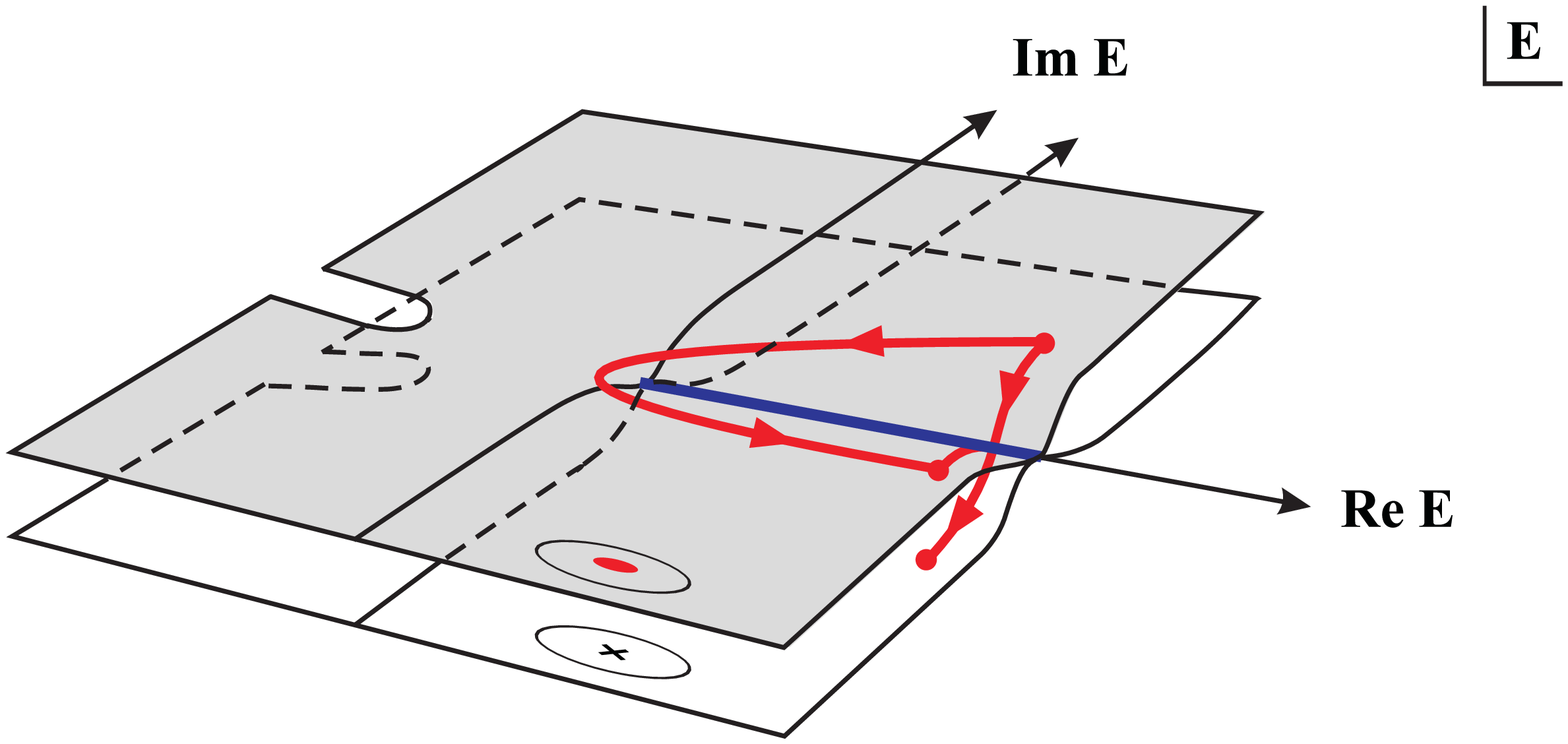}
\caption
{Illustration of the sheets and cut structure of the complex energy plane
in a world with just one channel. 
This represents the structure relevant to $\pi\pi$ scattering near its threshold. Experiment is performed on the top shaded sheet, just above the cut along the real energy axis. The {\it cross} on the lower, or second, sheet indicates where the $\sigma$-pole resides. The {\it ellipse} above this on the top, or first, sheet indicates where the $S$-matrix is zero. } 
\end{center}
\end{figure}
 Knowing the location of the pole, we can use the formalism sketched in Sect.~2 to determine its $\gamma\gamma$ couplings, since the
key $\pi\pi$ final state interactions automatically incorporate the effect of the $\sigma$. Dispersive computation then fixes the residue of the $\sigma$-pole in $\gamma\gamma\to\pi\pi$, and this corresponds to~\cite{mrp-prl}
\be
\Gamma (\sigma \to \gamma\gamma)\;=\;(4.1\,\pm\,0.3)\,\, {\rm keV}\quad .
\ee
\baselineskip=6.3mm
What does this mean for the composition of the $\sigma$? Simple predictions~\cite{babcock,barnes,achasov4q,barnesKK} are shown in Table 1. The
result in Eq.~4 seems to affirm a simple \uu, \dd composition.
Minkowski and Ochs~\cite{minkochs} have proposed that the $\sigma$ is largely a glueball. Perturbative arguments would suggest a small two photon coupling for such a state. This is certainly difficult to reconcile with a rate
of 4~keV.
However, the calculation of the $\qq$ composition is no more robust.
It can be checked by a relation provided by the naive quark model 
for states in the same $L$-band. Here, 
tensors and scalars of the same quark constitution, with the same spin ($S=1$) and the same orbital angular momentum ($L=1$), satisfy~\cite{chanowitz}:
\be
\frac{\Gamma(0^{++}\to\gamma\gamma)}{\Gamma(2^{++}\to\gamma\gamma)}\;=\;\frac{15}{4}\,\left(\frac{m_0}{m_2}\right)^n \quad ,
\ee
where the exponent $n$ in the mass term depends on the shape of the interquark potential at the relevant radius: $n=3$ for the short-distance Coulombic component, and $n\to 0$ for the linearly confining part. With the $\sigma$ and $f_2$ masses differing by more than a factor of 2, the result is very senstive to this exponent $n$. Moreover, Close {\it et al.}~\cite{rel-corr} have estimated that 
relativistic corrections can change Eq.~(5) by up to a factor of 2. More recently Lemmer~\cite{lemmer} has recomputed the radiative width of a $\KK$ molecule with alternative potential models and favours $\Gamma(f_0(980)\to\gamma\gamma) = 0.46^{+0.10}_{-0.13}$ keV, cf.~Table ~1. Once again we send out a plea for a genuine strong coupling calculation of the scalar radiative widths with different compositions
$\qq$, ${\overline{qq}}qq$ or glueball to compare with the result of 4~keV. Till then the  conclusion from Table~1 and Eq.~(4) of a $({\overline u}u +{\overline d}d)$ form for the $\sigma$ is perhaps too naive. 
\begin{table}[t]
\begin{center}
\vspace{-3mm}
\caption{Radiative widths of scalars in different models of their composition}
\label{table:3}
\renewcommand{\arraystretch}{1.3} % enlarge line spacing
\vspace{2mm}
\begin{tabular}{|c||c|c|c|c|}
\hline
\multicolumn{5}{|c|}{\rule[-0.3cm]{0cm}{5mm}{\large  $\Gamma(\gamma\gamma)$} keV }\\
\hline
composition  & (\uu + \dd)/$\sqrt{2}$ & \st & ${\overline{[ns]}}[ns]$, $\,n=(u,d)$ & $\KK$  \\
\hline
predictions & 4 keV & 0.2 keV & 0.27 keV & 0.6 keV \\
authors & Babcock \& Rosner~\cite{babcock}
& Barnes~\cite{barnes} & Achasov {\it et al.}~\cite{achasov4q} & Barnes~\cite{barnesKK}\\[2mm]
\hline
\end{tabular}\\
%\vspace{2.mm}
%\caption: Table 1: Two photon width for the scalars depending on their composition.
%\vspace{-2mm}
\end{center}
\end{table}

\newpage
\section{Amplitude analysis of $\gamma\gamma\to\pi\pi$}
\baselineskip=5.8mm

The combination of a low energy theorem and knowledge of $\pi\pi$ scattering means we can extract the $\gamma\gamma\to\pi\pi$ partial waves that sum to reproduce both the charged and neutral pion data of Fig.~3 near threshold. Now we need to separate the $S$-wave component of these cross-sections beyond 500 MeV, if we are to determine 
the two photon couplings of the many other $I=0$ scalars. 
Experimental data on $e^+e^-\to e^+e^-\pi\pi$ cover a limited angular range. This is because the identification of 
the charged pions (in particular) requires they have to be well-separated from the on-going $e^+$ and $e^-$ and distinguishable from the inevitably sizeable $\mu$-pair production background.  
Thus typically, data are restricted to $|\cos \theta^*| \le 0.6$. For neutral pions identified by their $\gamma\gamma$ 
decay mode, this separation is possible up to $\cos \theta^*=0.8$. In either case,  data cover only part of the 
angular range, and consequently, the partial waves are no longer orthogonal. This makes Amplitude Analysis impossible 
without additional information. Fortunately, this information  is provided by three facts:

\begin{itemize}
\item[1.] the low energy theorem allows the partial waves  to be determined below 500 MeV with both isospin 0 and 2 as described in Sects.~2 and 3,
\item[2.] the unitarity relation, Eq.~3, can be implemented if we have data
on the channels $H_n\to\pi\pi$. In practice, this extends up to energies where the two pion and $\KK$ channels saturate this relation. This is at most up to 
$ \sim 1400$ MeV, after which four pion and $\eta\eta$ channels can no longer be neglected, and
\item[3.] data exist on both $\pi^+\pi^-$ and $\pi^0\pi^0$ final states
to separate the $I=0,\ 2$ components, both of which are important because 
crossed-channel dynamics is dominated by  $\pi$ and $\rho$ exchange as discussed in Sect.~2.
\end{itemize} 
The procedure for how to do this was set out by David Morgan and myself~\cite{mp2} and applied to all $\gamma\gamma\to\pi\pi$ data available in 1999 by Elena Boglione and I~\cite{BP}. Then we found a number of solutions. The fit to the cross-sections is shown in Fig.~6 for the \lq\lq peak'' solution. Only the results integrated over the experimentally accessed range of $\cos \theta^*$ are shown. Of course the analysis fits the differential data too. Superimposed on this is the preliminary high statistics data from Belle on charged pion production~\cite{belle}. We see that these data (again both integrated and differential) provide results in 5 MeV bin steps that will inevitably alter
the determination of the underlying partial waves. Simple fits to the cross-section round the $f_0(980)$ using a Flatt\'e form have been performed by the Belle group to give a radiative width for this state of $(0.205\pm 0.089 \pm 0.132)$ keV~\cite{belle}. The large uncertainties come about because of the sizeable ambiguity in separating the signal from the background without the inputs 1,~2,~3 listed above. Once these data are finalised a new global Amplitude Analysis will be performed providing far more accurate (and perhaps different) results.
\begin{figure}[ht]
\begin{center}
\includegraphics[width=13.cm]{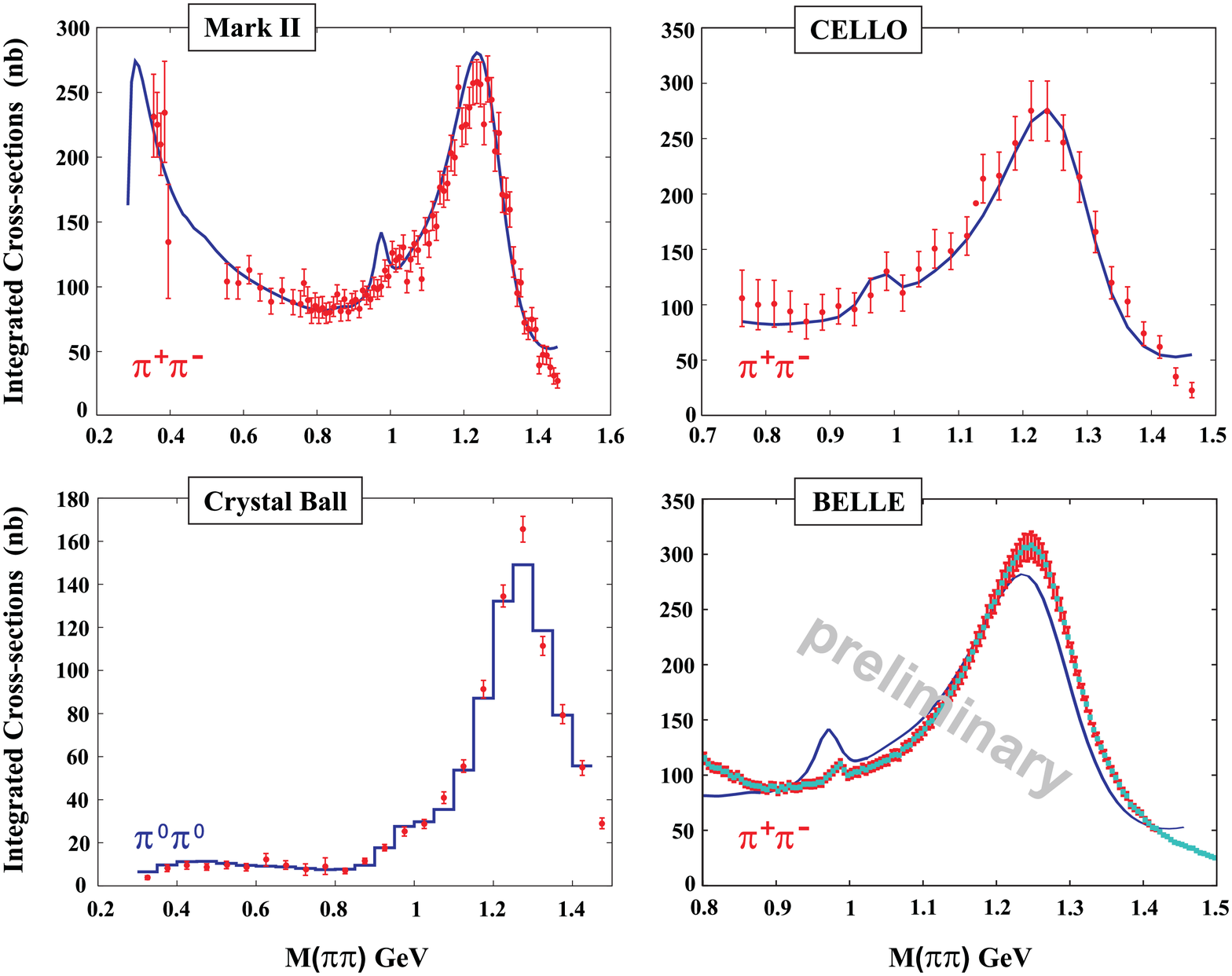}
%\vspace{2mm}
\caption
{The first three graphs show the integrated cross-section for 
$\gamma\gamma\to\pi\pi$ as a function of c.m. energy $M(\pi\pi)$ from 
Mark II~\protect\cite{MarkII}, Crystal Ball (CB)~\protect\cite{CB} 
and CELLO~\protect\cite{cello}.
The solid line defines the \lq\lq peak'' solution found in the 
Amplitude Analysis of Ref.~37. The fourth plot shows this same solution with 5 MeV resolution together with preliminary data from Belle~\protect\cite{belle} with its 
small statistical errors. The systematic uncertainties (not shown) are bigger.}
\end{center} 
\vspace{-5.mm}
\end{figure}

Comparing with the simple predictions in Table 1,
these are likely  to indicate that the $f_0(980)$ is not pure ${\overline s}s$, or ${\overline{[sn]}}[sn]$, or a $\KK$ molecule. In any event, these compositions are not orthogonal, but inevitably mix. In simple schemes, like that of  mixing the tetraquark and quarkonium nonets of Black {\it et al.}~\cite{schechter}, the two photon coupling of these states is a key ingredient in determining the components as discussed by Giacosa~\cite{giacosa}. 

The success of the naive quark model as a description of almost all hadrons comes about because mixing is usually weak. Consider as an example the well-known vector nonet. There the $\phi$, for instance, is thought of as an ideal ${\overline s}s$ candidate. In fact, it must have
 $\KK$ and $\rho\pi$  components too. It is through these that the physical $\phi$ decays. The dominant decays of the vector states are to pseudoscalars, {\it e.g.} $\rho\to\pi\pi$ and $\phi\to {\overline K}K$. These involve $P$-wave interactions which are naturally suppressed at threshold. This, together with the intrinsic strength of these couplings, makes the fraction of $\KK$ in the $\phi$ Fock space rather small, exactly as the $1/N_c$ expansion would imply. In contrast, as shown by~\cite{vanbeveren} Tornqvist, by van Beveren and collaborators, and by Boglione and myself, the situation is rather different for the scalars. The dynamics is controlled by $S$-wave thresholds, the generic role of which has been highlighted recently by Rosner~\cite{rosner} and by Jaffe~\cite{jaffedurham}. Even if the underlying scalar \lq\lq seed'' is purely \st, the resulting hadron  has a large $\KK$ component and its mass is forced close to 1 GeV. There is no natural suppression of quark loops for scalars.
 Meson loops inevitably mean
bare $\qq$ states have sizeable tetraquark components. The quark model and the observed hadrons are no longer aligned, and the differentiation displayed in Fig.~1 becomes artificial. To make sense of this once again requires genuine strong coupling predictions for the scalars, their masses, their lifetimes and their $\gamma\gamma$ couplings.

\section{Heavy flavour decays to light mesons}

Over the last few years,
the richest source of information on light hadron states has come from heavy flavour decays, especially from $e^+e^-$ colliders at charm and $B$-factories~\cite{mpreviews}.
This is because the major decay of $B$ and $D$-mesons is to $\pi$'s and $K$'s. For instance, the Dalitz plot for $D^0 \to K_s \pi^+\pi^-$  displays  marked two body structures in both $K\pi$ and $\pi\pi$ channels. These are generated by interferences with scalar states, that dominate the distribution. Indeed, in extracting the CKM angles from experimental data with minimal QCD input, one finds the biggest uncertainty in determining the angle  $\gamma$, or $\phi_3$, comes from the inaccuracies in treating the ubiquitous scalars. What is needed is a reliable description of the multitude of overlapping scalars discussed here. From these $B$ and $D$-decay data with their high statistics one should, by much more careful analyses, be able to map out the details of these scalars and consequently have precision information with which to uncover their structure, as well as better determine the CKM elements.

While we have focussed much of this discussion on the isoscalar $f_0$ states with their connection to Nambu's model of chiral symmetry breaking, one should not, of course, forget their isodoublet and isotriplet companions. Let us begin with the isodoublet $\kappa$. While the $K_0^*(1430)$ is well-established~\cite{pdg}, there has long been a controversy over whether there is a $\kappa$ at lower mass. LASS results~\cite{lass} on $K^-\pi^+$ scattering begin $\sim 200$~MeV above threshold at 825~MeV. Analytic continuation~\cite{cherry} of these data showed it to be highly unlikely there is a pole in the LASS energy range, corresponding to a $\kappa(900)$. However, the use of information on next-to-leading order Chiral Perturbation Theory on both $K\pi\to K\pi$ scattering and its crossed-channel of
$\pi\pi\to{\overline K}K$ in Roy-Steiner relations has enabled a $\kappa$-pole
to be located by Descotes-Genon {\it et al.}~\cite{descotes}. Indeed, further analysis~\cite{descotes2} has allowed a more precise continuation deep into the complex plane, much like that of Caprini {\it et al.} for the $\sigma$, using scattering data. This has accurately located the pole at  $E= 658 -i 278$ MeV with $\sim 13$ MeV uncertainties in both components. This position is at a much lower energy than simple Breit-Wigner fits to $D$-decay data from E791 and BES~\cite{kappaE791BES}. 
It is important to have checks on the key ingredients embodied in this continuation. This requires better experimental knowledge of $K\pi$ scattering amplitudes
down towards threshold. This will hopefully be provided by 
data on $D$-semileptonic decays, like $D^+\to K^-\pi^+ \mu^+\nu_{\mu}$, emerging from $B$-factories.

  Data from heavy meson decays and from $\gamma\gamma$ reactions are now accumulating significant statistics on the $\pi \eta$ channel. 
These pick out the isotriplet states, like the $a_0(980)$ and $a_0(1430)$. Particularly for the states close to ${\overline K}K$ threshold, this detail can bring new insights. To see how let us consider two simple scenarios, perhaps too simple to be realised in nature, in which the $f_0(980)$ and $a_0(980)$  are either intrinsically quark states or $K{\overline K}$ bound states, {\it i.e.} hadronic molecules~\cite{wi,achasovKK}. Telling the difference between these is 
like differentiating between whether the deuteron is intrinsically a six quark state, or is it a bound state of two baryons? The idea developed long ago by Weinberg~\cite{weinberg} is that a six quark state would be generated by short range forces, while a bound state of a proton and a neutron would be formed by longer range interactions. This is closely related 
to the question of whether a CDD pole is needed in an $N/D$ description of their amplitudes~\cite{cdd}: an argument Dalitz~\cite{dalitz} applied to the nature of the $\Lambda(1405)$. This consideration can elucidate mesons~\cite{morgan} like the $f_0$ and $a_0(980)$ too. Imagine another bifurcation of the sheets in Fig.~5 at a second threshold.
A state generated by short range interactions has a pole on two unphysical sheets, usually called the 2nd and 3rd sheets, while that generated by longer range $\KK$ forces only has a pole on the 2nd sheet~\cite{mpkk}. Precision mapping from data on the $f_0$ and $a_0(980)$ in the neighbourhood of $\KK$ threshold can in principle differentiate between these possibilities.  In reality, a state that is a mixture will have both poles, but how far that on the 3rd sheet is away is a measure of its $\KK$ component.
With the phenomenal body of data now becoming available on $B$, $D$ and $J/\psi$-decays, we should be able to do this mapping much more precisely than ever before and so really discover the composition of all  these intriguing scalars.
Perhaps, the Higgs sector of electroweak symmetry breaking will prove just as
rich.

\section*{Acknowledgments}
It is a pleasure to thank Professors Teijo Kunihiro, Makoto Oka, Takashi Nakano and Dr Daisuke Jido for the invitation to give this talk and for generous  travel funds from the Yukawa Institute.
The work presented here was partially supported by the EU-RTN Programme, 
Contract No. MRTN-CT-2006-035482, \lq\lq FLAVIAnet''.

\end{document}